\documentclass[12pt,preprint]{aastex}
\usepackage{graphicx}
\newcommand{\begen}{\begin{enumerate}}
\newcommand{\enen}{\end{enumerate}}
\newcommand{\beq}{\begin{equation}} 
\newcommand{\eeq}{\end{equation}} 
\newcommand{\beqa}{\begin{eqnarray}} 
\newcommand{\eeqa}{\end{eqnarray}} 
\newcommand{\p}{\partial}    
\newcommand{\pr}{^\prime}

\newcommand{\Msun}{M_{\sun}}
\newcommand{\Msunsec}{\Msun\,{\mathrm{sec}}^{-1}}
\newcommand{\cm}{{\mathrm{cm}}}

\newcommand{\erg}{{\mathrm{erg}}}
\newcommand{\second}{{\mathrm{s}}}

\newcommand{\cms}{\cm\,\second^{-1}}

\newcommand{\gcc}{\mathrm{g}\,\cm^{-3}}

\def\ltaprx {\lower .1ex\hbox{\rlap{\raise .6ex\hbox{\hskip .3ex
        {\ifmmode{\scriptscriptstyle <}\else 
                {$\scriptscriptstyle <$}\fi}}}
        \kern -.4ex{\ifmmode{\scriptscriptstyle \sim}\else 
                {$\scriptscriptstyle\sim$}\fi}}}
\def\gtaprx {\lower .1ex\hbox{\rlap{\raise .6ex\hbox{\hskip .3ex
        {\ifmmode{\scriptscriptstyle >}\else 
                {$\scriptscriptstyle >$}\fi}}}
        \kern -.4ex{\ifmmode{\scriptscriptstyle \sim}\else 
                {$\scriptscriptstyle\sim$}\fi}}}

\begin{document}

\title{Nucleosynthesis in Outflows from the Inner Regions of Collapsars}
\author{Jason Pruet\altaffilmark{1}, Todd A.~Thompson\altaffilmark{2,3} \&
R.~D.~Hoffman\altaffilmark{1}}
\altaffiltext{1}{N-Division, Lawrence Livermore National Laboratory, Livermore CA 94550; pruet1@llnl.gov,rdhoffman@llnl.gov}
\altaffiltext{2}{Hubble Fellow}
\altaffiltext{3}{Astronomy Department 
and Theoretical Astrophysics Center, 601 Campbell Hall, 
The University of California, Berkeley, CA 94720; 
thomp@astro.berkeley.edu}

\begin{abstract}

We consider nucleosynthesis in outflows originating from the inner regions of
viscous accretion disks formed after the collapse of a rotating massive
star. We show that wind-like outflows driven by viscous and neutrino heating
can efficiently synthesize Fe-group elements moving at near-relativistic
velocities. The mass of $^{56}$Ni synthesized and the asymptotic velocities 
attained in our calculations are in accord with those inferred from observations 
of SN1998bw and SN2003dh. These steady wind-like outflows are generally proton rich,
characterized by only modest entropies, and consequently synthesize essentially nothing 
heavier than the Fe-group elements. 
We also discuss bubble-like outflows resulting from rapid
energy deposition in localized regions near or in the accretion disk. 
These intermittent ejecta emerge with low electron fraction and are a 
promising site for the synthesis of the ${\rm A=130}$ $r$-process peak elements.

\end{abstract}

\keywords{gamma rays: bursts---nucleosynthesis---accretion disks}

\section{Introduction}

In this paper we examine the production of nuclei in matter escaping the
innermost regions of collapsars. Collapsars occur when the 
usual neutrino-powered 
supernova shock fails to expel the mantle
of a rotating massive star whose core has collapsed \citep{woo93,macfadyen}.
The inner parts of the star collapse to form a disk accreting rapidly
onto a central black hole. Interesting elements, and in particular
$^{56}{\rm Ni}$ and ${\rm A=130}$ peak $r$-process elements, may be synthesized
in outflows from the inner regions of this disk. 

Understanding how collapsars make $^{56}{\rm Ni}$, whose decay fuels
optical light curves of SNe, is central to connecting the deaths
of massive stars with Gamma Ray Bursts (GRBs). There are now observations
of SN-like light curves for some five or six GRBs \citep{pri03}. For two
SNe/GRBs (SN 1998bw--\cite{gal98} and  SN2003dh--\cite{sta03,hjo03})
there are detailed estimates of Ni ejecta mass
and velocity (\cite{pat01,iwa98,woo99} for SN1998bw and  
\cite{hjo03,woo03} for SN2003dh).

Though it is generally agreed that collapsars are promising sources
for the observed Ni, details remain uncertain. There
are several possibilities. As in ``ordinary'' SNe, Ni may be
synthesized explosively as a strong shock traverses the stellar
mantle. Parametrized piston-driven simulations of the explosion of
massive stars \citep{woo03} and simulations of massive stars exploded
by outgoing bi-polar jets \citep{mae03} show that this mechanism may
produce substantial amounts of fast moving ${\rm Ni}$. It is not
clear, though, if explosive burning can account for the very fast
($v\gtrsim 0.1c$) and massive ($M_{\rm Ni}\approx 0.5\Msun$) outflows
seen in SN2003dh. Another possibility is that the Ni is synthesized
in some sort of slow, heavily baryon polluted outflow formed above the
black hole \citep{nag03}. The last possibility, investigated here, is
that Ni is synthesized in wind-like outflows from the accretion disk
\citep{macfadyen, mac03}. 
Those authors find, for plausible disk viscosities and
accretion rates, fast ($v\sim 4\times 10^9 \cms$) Ni-rich outflows
blown off of the inner accretion disk by viscous heating. We provide a
simple treatment of these winds with an eye toward understanding the wind
energetics and the 
conditions needed for efficient Ni synthesis.

We also discuss the possibility of synthesizing $r$-process elements 
in collapsar events.  The $r$-process (Burbidge et al.~1957; Cameron 1957) 
accounts for roughly half the heavy nuclides above the iron group, producing
characteristic abundance peaks at $A\sim80$, 130, and 195.  The
astrophysical site for the production of these elements remains uncertain.
By examining the relative isotopic abundances of $^{182}$Hf/$^{180}$Hf 
and $^{129}$I/$^{127}$I, Wasserburg, Busso, \& Gallino (1996) argued that 
at least two distinct $r$-process sites, operating in the galaxy with different 
rates, must produce the nuclides with $A\lesssim130$ and $A\gtrsim130$ (see 
also Wasserburg \& Qian 2000).  Subsequent work by Qian, Vogel, \& Wasserburg (1998) 
and Qian \& Wasserburg (2000) argues that events which produce the heavy, third-peak 
nuclei occur ten times more frequently than events which produce the lighter 
(second peak and below) elements.  The latter must also produce iron-group
elements copiously. Conversely, the high frequency (producing $A\sim195$)
must produce little iron.  The association of lower event frequency with
large Ni production, plus the work of Heger et al.~(2003), which argues
that the collapsar rate might be $\sim10$\% of the total core-collapse
supernova rate,  points potentially to collapsar events as the 
astrophysical site for $A\lesssim130$ $r$-process element synthesis.
We argue that the inner collapsar disk material 
with low electron fraction may be ejected rapidly, with modest entropy, 
in magnetically dominated filaments or bubbles and that this 
intermittent outflow is a likely site for $r$-process nucleosynthesis up to $A\sim130$.

\section{Disk Outflows and Nucleosynthesis}

Nucleosynthesis in outflows from the disk is primarily sensitive to
the electron fraction ($Y_e$), the entropy per baryon ($s/k_b$), and
the timescale characterizing the expansion of the fluid around
the time of efficient $\alpha$-particle formation \citep{hof97}. To
estimate these quantities we examine two different, and in some sense
limiting, realizations of the outflow. The first is a hydrodynamic picture of
the outflow, which assumes the presence of stable and ordered pressure
profiles: a steady disk wind. The second picture we examine is one where rapid 
magnetic reconnection or turbulent viscous heating deposits entropy (and energy) in 
localized ``bubbles'' within the disk, causing rapid ejection of low $Y_e$ material. 
Both types of processes likely occur to some extent. 

\section{A Hydrodynamic Wind Picture}

Steady or quasi-steady spherical winds have been extensively used to 
study nucleosynthesis in the neutrino-driven wind occurring several
seconds after core bounce in core-collapse SNe 
(Duncan, Shapiro, \& Wasserman 1986; Woosley et al.~1994; Takahashi, Witti, \& Janka 1994;
Qian \& Woosley 1996; Cardall \& Fuller 1997; Sumiyoshi et al.~2000;
Otsuki et al.~2000; Wanajo et al.~2001; Thompson, Burrows, \& Meyer~2001).
Of particular interest to the present study is the paper by Qian \& Woosley (1996), 
which provided insight into and analytic expressions for describing the connection between the various
parameters determining the wind itself (neutron star radius and mass, neutrino luminosities, etc.)
and the properties of the flow that determine the resulting nucleosynthesis: the electron fraction,
entropy, and dynamical timescale.  It should be noted that so far no
agreed upon neutron star wind solutions give a robust $r$-process (with the possible
exception of winds from very relativistic neutron stars \citep{cardall,thompson1}
or highly magnetic neutron stars (Thompson 2003)).

In this section we describe our simple model of time-independent disk winds, akin
to the models previously developed in the neutron star context.
Roughly, our assumption of a steady-state outflow is an attempt to use an estimate
of the vertical disk pressure gradient to derive the dynamical timescale of the
outflow and the increase in entropy due to viscous and neutrino energy
deposition.  Because we parameterize flow streamlines and currently employ
only a local $\alpha$-disk prescription for energy deposition, our analysis
is limited and must be tested eventually against full MHD simulations of collapsar
disk winds.  

\subsection{Equations}

The equations governing the wind structure are determined by mass, momentum,
energy, and lepton number conservation. For steady flow these are,
respectively,
\beqa
\label{bnc} \nabla \cdot (\rho u) &=& 0 \\
\label{momcon} \rho u \cdot \nabla u &=& \vec{f} \\
\label{econ} u \cdot (\nabla \cdot T)&=&-u\cdot(\nabla \cdot T_{\alpha})+q_{\nu} \rho N_a \\
\label{yecon} u\cdot (\nabla Y_e) &=& -\lambda_{e^-{\rm p}}\left(Y_e -\frac{1-X_{\rm free}}{2}\right) +
\lambda_{e^+{\rm n}}\left(1-Y_e-\frac{1-X_{\rm free}}{2}\right).
\eeqa
Here $\rho$ is the mass density, $u$ is the velocity, $q_{\nu}$ is the
net neutrino energy deposition rate per baryon, and $\vec{f}$ is the
force (gravitational+viscous+pressure gradient) acting on outflowing
fluid elements.  $T=\rho u\otimes u+pg$ is the stress-energy tensor of
the fluid, with $p$ the pressure and $g$ the metric tensor. Our
treatment is entirely Newtonian, and in spherical polar coordinates
($1=r$, $2=\theta$, $3=\phi$) $g_{11}=1$, $g_{22}=r^2$,
$g_{33}=r^2\sin^2\theta$, and $g_{ij}=0$ for $i\neq j$. Including relativistic
effects in aspherical viscous flows is straightforward, but more complicated
than including relativistic effects for inviscid spherical outflows
from neutron stars. 

In eq.~(\ref{yecon}) we have approximated the flow as consisting of free nucleons and $\alpha$-particles,
$Y_e$ is the net number of protons per baryon in the flow, $\lambda_{e^-{\rm p}}$ and 
$\lambda_{e^+{\rm n}}$ are the rates for electron and positron capture
on free nucleons (e.g.~Qian \& Woosley 1996),  and $X_{\rm free}$ is the mass fraction
of free nucleons in the flow. Lepton capture on heavy nuclei is typically very slow compared
to lepton capture on free nucleons and is neglected here. When $X_{\rm free}=0$ (all $\alpha$-particles), 
then, eq.~(\ref{yecon}) gives zero rate of change for $Y_e$.
We approximate $X_{\rm free}$ as
\beq
\label{xfree}
X_{\rm free}=8.2\cdot 10^8\frac{T_{\rm MeV}^{9/8}}{\rho ^{3/4}} \exp\left(\frac{-7.074}{T_{\rm MeV}}\right)
\eeq
or unity, whichever is smaller \citep{woosbar}. Here $T_{\rm MeV}$ is the temperature in 
MeV. Eq.~(\ref{yecon}) does not account for the 
influence of neutrino capture on $Y_e$. Including neutrinos in a 
reliable way is beyond the scope of this paper.
However, in section \ref{yesection} we outline in general terms how
neutrinos are expected to influence $Y_e$. 

The viscous stress tensor is (e.g. \cite{mihxx}) 
\beq 
T_\alpha=2\mu D,
\eeq 
with 
\beq
D^{xy}=\frac{1}{2}\left(u^{x\,y}_{\,\,\,;}+u^{y\,x}_{\,\,\,;}\right).
\eeq 
Here a semi-colon represents the covariant derivative.
Following \cite{sto99}, we neglect all components of $D$ except
$D^{r\phi}=D^{\phi r}=(1/2)u^\phi_{,r}$ and
$D^{\theta\phi}=D^{\phi\theta}=(1/2r^2)u^\phi_{,\theta}.$ The neglect
of the other components of the stress tensor is justified if the
magnetic instabilities providing the shear stresses produce small
poloidal stresses. For the coefficient of shear viscosity we adopt the
parametrization often used in studies of viscous disks: $\mu=\alpha
p/\Omega_k(r_0)$. Here $\alpha$ is the standard disk alpha parameter
\citep{sun50}, $r_0$ is the radius from which the wind leaves the
disk, and the Keplerian frequency is $\Omega_k(r_0)=\sqrt{GM/r_0^3}$,
with $M$ the mass of the central black hole.

To proceed, we parametrize the trajectories of the outflow by
$\vec{r}=(r,\theta,\phi)=(r_0g(\tilde \theta),\tilde \theta,\phi)$ and the wind
velocity by $\vec{u}=(\dot{r},\dot\theta,\dot\phi)=(r_0g\pr f, f,
\xi)\Omega_{k}(r_0)$.  Throughout the paper a prime denotes differentiation
with respect to $\tilde \theta$. We also define the velocity in
the $r-\theta$ plane, $v=f\Omega_{k}(r_0)\sqrt{g^2+g^{\prime 2}}$.
The basic idea of parametrizing the trajectories in the way we have comes
from the pioneering work of \cite{bla77}. Unlike that work, however,
we are not solving for the outflow and magnetic field configuration in
a global and self consistent way.
Rather, we will choose trajectories $g(\tilde \theta)$ in
order to get estimates of the wind parameters important in determining
nucleosynthesis. Solving for the trajectories in a consistent way
likely requires a hydromagnetic simulation which can capture the
interplay between the outflow and the magnetic field configuration.

With the above definition for the streamlines, eq.~(\ref{bnc}) becomes
\beq
\label{aequation}
\rho v a = {\rm Constant}.
\eeq
Here $a=g^3 \sin\theta (g^2+g^{\prime 2})^{-1/2}$ and is proportional to the
area defined by fluid streamlines. 
The equation governing $v$ is found
by taking the projection of eq.~(\ref{momcon}) along the streamlines:
\beq
vv\pr\left(1-\frac{P_{,\rho}}{v^2}\right)=A_v-\frac{P_{,s}s\pr}{\rho}+P_{,\rho}\frac{a\pr}{a},
\eeq
where $P_{,\rho}=\p P/\p\rho|_s=c_s^2$, $P_{,s}=\p P/\p s|_\rho$, and
\beq
A_v=r_0^2\Omega_{k}^2\xi^2\left(\frac{1}{2}g^2\sin^2\theta\right)\pr+r_0^2\Omega_{k}^2\left(\frac{1}{g}\right)\pr
\eeq
is the sum of the gravitational and centrifugal forces. 
Eq.~(\ref{econ}) can be recast as
the equation governing the entropy of the outflow
\beq
\label{sdot}
f\Omega_{k}s\pr=\frac{\alpha P \sin^2 \theta \Omega_k(r_0)}{\rho T N_a}\left(\left(\frac{3\xi}{2}\right)^2+\left(\xi\pr+\frac{3g\pr \xi}{2g}\right)^2\right)
+\frac{q_\nu}{T}.
\eeq
Here $s$ is the entropy per baryon scaled by $k_b$. 
We approximate $s$ as the sum of contributions from 
relativistic light particles (photons and $e^{\pm}$ pairs), free nucleons, and alpha
particles:
\beq
\label{sequation}
s=5.21\frac{T_{\rm MeV}^3}{\rho_8}+X_{\rm free}\left(12+\ln\left(\frac{T_{\rm MeV}^{3/2}}{\rho_8}\right)\right)
+\frac{1-X_{\rm free}}{4}\left(15.4+\ln\left(\frac{T_{\rm MeV}^{3/2}}{\rho_8}\right)\right),
\eeq
where $\rho_8=\rho/10^8{\gcc}$.
The influence of nuclear recombination on the wind dynamics is discussed in section \ref{nucrecomb}.
 The equation describing the evolution
of $Y_e$ is
\beq
f\Omega_{k} Y_e\pr= 
 -\lambda_{e^-{\rm p}}\left(Y_e -\frac{1-X_{\rm free}}{2}\right) +
\lambda_{e^+{\rm n}}\left(1-Y_e-\frac{1-X_{\rm free}}{2}\right).
\label{yecon2}
\eeq

We adopt a crude parametrization for $q_{\nu}$:
\beq
\label{qnu}
q_\nu=q_h-q_c.
\eeq
 Neutrino energy loss
from the wind occurs principally via $e^{\pm}$ capture on free
nucleons ($e^-p\rightarrow n\nu_e$ and $e^+ n \rightarrow p \bar{\nu}_e$). 
The energy loss rate associated with these processes is
$q_c\approx 2.3 T_{\rm MeV}^6 ({\rm MeV/sec\, baryon})$.
As in the neutrino-driven
winds that occur in the late-time core-collapse SN cooling epoch, neutrino heating occurs
principally via charged-current neutrino capture on free nucleons
($\nu_e n \rightarrow p e^-$ and $\bar{\nu}_e p\rightarrow n e^+$). The
heating rate for these processes is $q_h\approx
5L_{\nu,51}(\langle{E_{\nu}}\rangle /10 {\rm MeV})^2 (1/r_7^2)({\rm MeV/sec\,baryon})$ 
\citep{qia96}. Here we have approximated the inner, neutrino luminous
portions of the disk as being spherical, $L_{\nu,51}$ is the sum of the $\nu_e$ and 
$\bar{\nu_e}$ neutrino luminosities in units of $10^{51}{\rm erg/sec}$, $\epsilon_{\nu}$
is the average $\nu_e$ or $\bar{\nu_e}$ energy, and $r_7=r/10^7{\cm}$. 

To get an idea of the relative importance of viscous and neutrino
heating, note that for a 3$\Msun$ black hole with Kerr parameter $a=0$ 
accreting at
$0.1\Msunsec$, Popham, Woosley, \& Fryer~(1999) estimate 
$L_{\nu,51}\approx 3-7$, for $\alpha$ in the range $0.03-0.1$. Typical average
neutrino energies for these disks are $\langle E_{\nu} \rangle \approx 15-20
{\rm MeV}$.  At $r=10^7{\rm cm}$, the viscous
heating rate is $\approx 140T_{\rm MeV}(\alpha/0.1)({\rm MeV/sec\,
nucleus})$ for $\xi\approx 1$, while $q_h\approx
100(L_{\nu,51}/5)(\epsilon_{\nu}/20{\rm MeV})^2({\rm MeV/sec\,
nucleon})$. For these disks, then, neutrino and viscous heating can be
comparable.  For a disk surrounding a black hole with high Kerr parameter,
neutrino heating can dominate over viscous heating. Popham, Woosley, \& Fryer~(1999) estimate that a disk with $a=0.95$ has a neutrino luminosity about
eight times larger than the same disk with $a=0$. As we discuss though, 
the parameters important for nucleosynthesis are not very dependent on the
neutrino luminosity (though the influence of neutrino losses is important
in determining the disk structure).

\subsection{Wind Profiles: $Y_e$, $s$, and $\tau_{\rm dyn}$}

With the above formalism we can discuss conditions in outflows 
from the disk and implications for nucleosynthesis. 
For simplicity, fluid streamlines are taken
to be straight lines making an angle $\theta_0$ with the plane of the
disk (or $\pi-\theta_0$ with the z-axis). We assume that $\xi$
decreases with distance $z$ above the disk as
\beq
\xi=\exp(-z/\xi_z)
\eeq
with $\xi_z$ the scale height for the decrease in $\xi$. 
Thus, as the wind moves out of the plane of the disk, its velocity in 
the $\phi$ direction evolves as $\xi(z)\Omega_k$. Our discussion and 
parametrization of these steady state winds is similar in some ways to the work
of \cite{dai02}, who examined the conditions needed for ultra-relativistic
(Lorentz factor much larger than unity) outflows from accretion disks.

We do not present an exhaustive survey of wind models -- the simplicity
of our model probably does not warrant it. Instead we outline how the
parameters influencing nucleosynthesis in the wind depend on the
character of the accretion disk and on the starting radius of the
outflow.  In Table \ref{tbl1} we show results from wind solutions for
outflows from a moderate viscosity ($\alpha=0.1$) disk and from a low
viscosity ($\alpha=0.03$) disk. In both cases the accretion rate of
the disk is $\dot M=0.1\Msunsec$. For each type of disk, outflows from
a moderate radius ($r_0=10^7$) and a small radius ($r_0=10^{6.5}\cm$)
are considered. Initial (in-disk) values of the temperature and
density for the calculations were taken from the results presented in
\citet{pop99}.  Typical disk temperatures are a few MeV, typical
densities are $\rho\gtrsim 10^9\gcc$, and typical entropies are of
order 5-10.  The starting electron fraction was taken from
\citet{pru03}. The electron fraction in the disk depends sensitively
on the mass accretion rate and viscosity, and can be anywhere in the
range $0.1\lesssim Y_e\lesssim 0.53$. Results shown in Table
\ref{tbl1} were calculated with a neutrino heating rate
$q_h=(100/r_7^2)({\rm MeV/sec \,nucleus})$. We find that changing $q_h$ by a factor
of two in either direction does not have a big influence on the asymptotic
wind parameters. Increasing $q_h$ by a factor of eight results in an increase
in the mass outflow rate of $\approx 25\%$ and in increase in the asymptotic
entropy of $\approx 3$ units for models C and D. 
All calculations in Table \ref{tbl1} are for
$\theta_0=80\degr$. Effects of changing $\theta_0$ are discussed in
\S\ref{flung}.

\begin{figure}
\epsscale{0.8}
\plotone{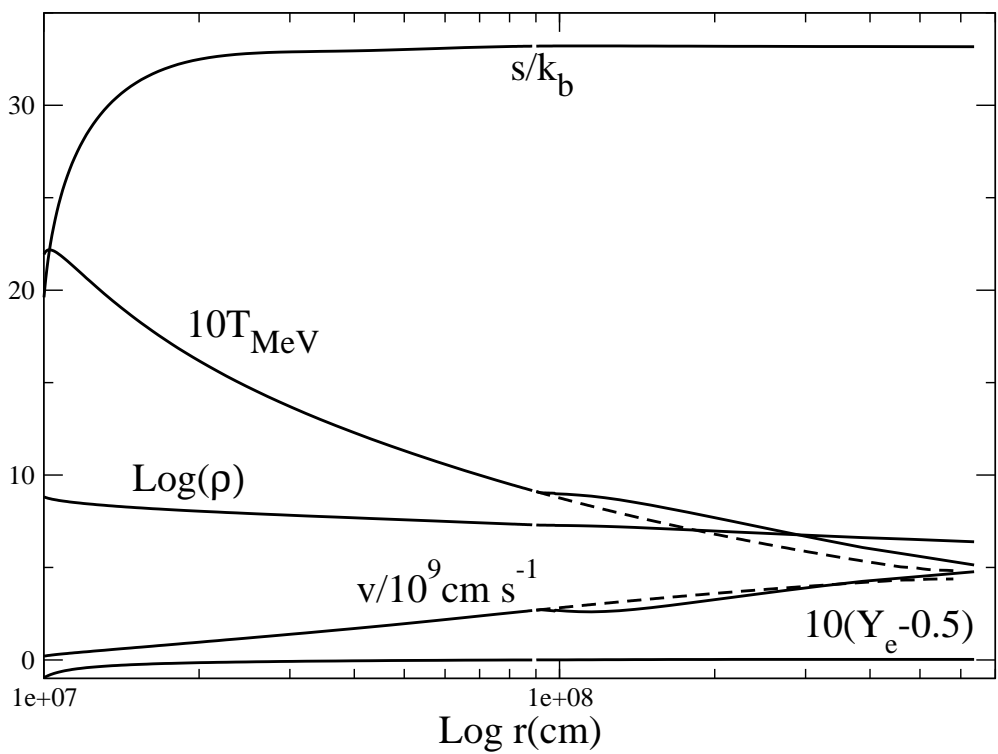}
\caption{Wind for material beginning at $r_0=10^7{\rm cm}$ from 
a disk with $\alpha=0.1$ and $\dot M=0.1\Msunsec$. The solid lines are
for calculations including $\alpha-$recombination, while the dashed
lines are for calculations which neglect the influence of $\alpha$ recombination on the wind.
Note that in our simple calculations most of the heating (entropy change) occurs near the base of
the flow. \label{a}}
\end{figure}

\begin{figure}
\epsscale{0.8}
\plotone{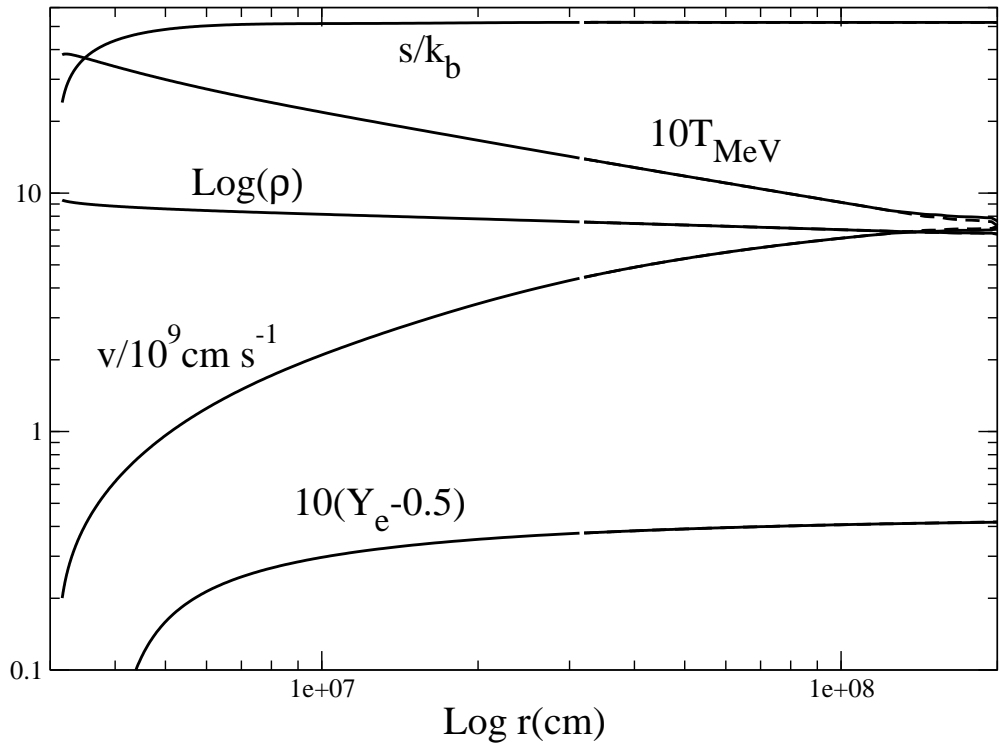}
\caption{The same as Fig. \ref{a} except for material beginning at
$r_0=10^{6.5}\cm$. Note that this figure has logarithmic spacing on the
vertical axis.
\label{b}}
\end{figure}

To give a point of reference for the following discussion we show
typical wind solutions in Figures \ref{a} and \ref{b}.  Figure 1
corresponds to a wind beginning at $r_0=10^7\cm$ in a disk with
$\alpha=0.1$ accreting at a rate of $0.1\Msunsec$ onto a central black
hole of mass $3\Msun$. These parameters are close to those thought to
describe conditions in collapsars. For the calculation in
Fig. \ref{a}, $\theta_0=80\degr$ and $\xi_z=2r_0=2\times10^7\cm$. Figure
\ref{b} shows a wind for the same parameters as in Fig. \ref{a} except
with $r_0=\xi_z/2=10^{6.5}\cm$. These winds bear qualitative
similarities to $\nu-$driven winds from neutron stars. The temperature
at the base of the wind is approximately that for which the heating
and cooling rates balance each other. Also, most of the heating occurs
at the base of the flow, with the evolution at larger radii being isentropic.

\subsubsection{The Asymptotic Entropy}
\label{entsection}
The trend of greater increase in final entropy with increasing initial
gravitational potential and the weak dependence of the final entropy
on the heating rate ($\alpha$) is evident. This is similar to the case
for winds from NS's. \citet{qia96} argued that the final entropy
should scale as $\sim r_0^{-2/3}$, with only a weak dependence on the 
heating rate, which is a fair
approximation to the results shown in Table \ref{tbl1}. Overall, the
final entropies expected for winds from the disk are rather modest
($\sim 30-50$) and more typical of the $\alpha$-process than the
$r$-process.

So far it is not clear to what extent the wind properties relevant
for nucleosynthesis are determined by our simplified model
and to what extent the wind properties are determined by more basic 
considerations. To get insight into the distinction note that the hydrodynamic
equations can be recast in the form
\beq
\label{bigbern}
Q\pr\equiv Ts\pr=b\pr
\eeq
with $Q$ the total energy per baryon added to the flow by viscous
and neutrino heating and $b$ the Bernoulli integral for this system:
\beqa
\label{bern}
b&\equiv& \frac{m_N v^2}{2}+Ts_{rad}+\frac{5T}{2\bar{A}}+\epsilon_{\rm nuc}-\frac{GMm_N}{r}(1+U_c)\\
&\approx& 500\left( \frac{v}{c} \right)^2 + T_{\rm MeV} s_{\rm rad}
+\frac{5T_{\rm MeV}}{2\bar{A}} +\epsilon_{\rm nuc}
-45(1+U_c)
\left( \frac{M}{3\Msun} \right) \frac{10^7\cm}{r} ({\rm MeV}).
\eeqa
Here $s_{rad}$ is the entropy per baryon in radiation, $\bar A$
is the mean atomic mass of bound nuclei and $\epsilon_{\rm nuc}$ is the mean 
nucleon binding energy. If everything burns to $\alpha$-particles 
the difference in $\epsilon_{\rm nuc}$ before and after nuclear recombination 
is about 7.074MeV, while if everything burns to Fe-group nuclei the difference
is about 1.4 MeV larger. The quantity $U_c$ accounts for the influence of the
$\phi$ component of the velocity in overcoming the gravitational potential. 
For example, within the disk $r_0\dot\phi\approx r_0\Omega_k(r_0)$, 
so the material already has half the kinetic energy needed to escape 
the gravitational pull of the black hole. In our treatment 
\beq
\label{uceq}
U_c=(g/2)\int_{\pi/2}^{\tilde \theta} \xi^2 (g^2 \sin^2 \theta)\pr d\theta.
\eeq 

The change in entropy of the outflowing fluid is 
\beq
\label{deltas}
\Delta s \equiv s_f-s_i=\int_{i}^{f}\frac{db}{T},
\eeq
where the subscripts $i$ and $f$ denote values at the base and end 
(i.e. asymptote) of the flow respectively. If the temperature decreases
along outflowing streamlines
\beq
\label{deltas2}
\Delta s \gtrsim \Delta b/T_i.
\eeq
Approximate 
equality in eq.~(\ref{deltas2}) holds if most of the heating occurs very
near the base of the flow. 

For a given starting radius and disk composition the above
considerations give a minimum value for the final entropy as a
function of the asymptotic velocity of the outflow and $U_c$. Values for the
different cases considered in the wind calculations are shown in Table
\ref{tbl3}. The calculations there assume $U_c=0$ (or equivalently that 
$\xi_z$ is small). In this case the centrifugal
potential plays no role. 
There is fairly close agreement between the 
asymptotic entropies found from consideration of the Bernoulli
integral and the asymptotic entropies found in our wind calculations
(Table \ref{tbl1}). Entropies found in our calculations are typically
about 5 units lower than the values in Table \ref{tbl3}. 
This arises mostly because our wind calculations have modest
$\xi_z$ and the rotational velocity plays some role 
in decreasing the effective gravitational potential.

\subsubsection{The Asymptotic Electron Fraction}
\label{yesection}
In contrast to the way in which the final entropy is set, the final
electron fraction in these disk winds is set by quite different
factors than in NS winds. In winds from NS's, neutrinos dominate both
the energy deposition rates and the lepton capture rates. The neutron
to proton ratio comes into approximate equilibrium with the neutrino
spectra. Because the neutrinos originate from the neutron rich crust
of the neutron star, which has a high opacity to electron neutrinos,
the $\bar{\nu}_e$ spectrum is hotter than the $\nu_e$
spectrum. Consequently, neutrino capture above the nascent neutron
star leads to a neutron-rich wind favorable for the $r-$process.

In winds from accretion disks that are optically thin to neutrinos
all factors conspire to make $Y_e>0.5$. In the first place,
$e^{\pm}$ capture, rather than neutrino capture, generally sets 
$Y_e$ in the disk and in the wind. 
As viscous heating adds entropy to the outgoing fluid 
the electron degeneracy is removed. Weak equilibrium then favors $Y_e>0.5$
because of the neutron-proton mass difference.

Secondly, when neutrino
captures are important they tend to increase $Y_e$ \citep{sur03,bel03}. 
Very roughly, this can be
thought of as a consequence of the neutrinos carrying net lepton number away
from the deleptonizing disk. To make these arguments more quantitative, note 
that the ratio of the rates for $\nu_e$ and $\bar{\nu}_e$ capture is
\beq
\label{rnu}
R\equiv
\frac{\lambda(\bar{\nu}_e{\rm p}\rightarrow e^+{\rm n})}
{\lambda({\nu}_e{\rm n}\rightarrow e^-{\rm p})}=
\frac{{n}_{\bar{\nu}_e}}{n_{{\nu}_e}}
\left(\frac{1.2\langle E_{\bar{\nu}_e} \rangle^2
+\Delta^2 -2\Delta \langle E_{\bar{\nu}_e} \rangle} {1.2\langle
E_{{\nu}_e} \rangle^2 +\Delta^2 +2\Delta \langle E_{{\nu}_e}
\rangle} \right)
\eeq 
\citep{qia96}. 
In eq.~(\ref{rnu}) $\Delta=1.293{\rm
MeV}$ is the neutron-proton mass difference, $\langle E_{\nu} \rangle$
is the average neutrino energy, and $n_{\nu}$ is the neutrino number
density. The factor 1.2 weighting $\langle E_{\nu} \rangle^2$ is
approximate and depends on details of the neutrino spectrum.  Taking
typical neutrino emission as originating from $r\approx 10^{6.5}{\rm
cm}$ in the disk, the disk parameters from \citep{pop99} and the electron 
fraction from \citep{pru03}, the ratio in eq.~(\ref{rnu}) 
can be calculated \citep{ffn}. We find $R=0.94$ for the $\dot
M=0.1\Msunsec,\alpha=0.1$ disk and $R=0.66$ for the $\dot
M=0.1\Msunsec,\alpha=0.03$ disk. Because $R<1$, neutrino capture above
the disk tends to drive the outflow proton rich. 

It is worth noting
that in the very inner regions of the disk the composition can fall
out of weak equilibrium \citep{pru03}, with $Y_e$ smaller than the equilibrium
electron fraction $Y_{e,eq}$. This results in a low electron Fermi
energy, and consequently, relatively underluminous ${\nu}_e$
emission. For example: for the $\alpha=0.03$ disk, material within the disk 
at $r=10^{6.2}$ cm
has $Y_e/Y_{e,eq}\approx 0.7$ and neutrinos originating from this region
are characterized by $R=1.2$.  By themselves, neutrinos from a region
like this would tend to drive the outflow {\it neutron} rich. However,
relativistic effects quash the influence of neutrinos from the
innermost regions of the disk. The neutrino capture rate far above the
disk goes approximately as the fifth power of the redshift factor
$h=\sqrt{1-2M/r_0}$ (e.g. \cite{pru00}). This reduces the influence
of neutrinos from $r_0=10^{6.2}\cm$ by about an order of magnitude
relative to the influence of neutrinos from $r=10^{6.5}{\rm cm}$.  For
disks accreting more rapidly than $\approx 1\Msunsec$, the inner
regions become opaque to neutrinos and a more careful treatment of the
neutrino spectrum is needed \citep{sur03}.

The above considerations about $Y_e$ are exemplified in Table \ref{tbl1}.
In all cases the asymptotic electron fraction is larger than the in-disk
electron fraction. The scaling of final $Y_e$ with $r_0$, $\xi$, and $\alpha$
has clear origins. All else being equal, a smaller $\alpha$ implies 
a denser disk, with faster weak interaction rates. In addition, a higher 
density implies - for a given mass loss rate - a lower outflow
velocity and more time for weak processes to operate. These are the reasons
why $Y_e$ is so large in the $\alpha=0.03$ disk. Similar reasons are 
behind the scaling of $Y_e$ with $r_0$. A smaller $r_0$ implies a 
larger density. As well, the entropy of the outflow (and the relative 
importance of positron capture) increases with decreasing $r_0$. Lastly, 
as $\xi_z$ increases, the wind material is flung out centrifugally, attaining 
larger velocities at smaller distances from the disk, and there is less time 
for positron capture. This is why the asymptotic
electron fraction in the $\xi_z=4r_0$ case is relatively low. 

Though disks with low $\alpha$ (or high $\dot M$) can be very neutron rich
(\cite{pru03,bel03}), wind-like outflows will not preserve the neutron
excess. In particular, wind-like outflows as discussed here cannot result
in an asymptotic $Y_e\lesssim0.4$. An exception to this is for the relativistic
jet originating very near the black hole. Neutrino-antineutrino annihilation
and relativistic effects dominate such ultra-relativistic 
outflows and they can remain
neutron rich \citep{pru00}.

It should be noted that we may overestimate the electron-fraction in 
the $\alpha=0.1$ disk. This is because our calculation shows about
half of the change in $Y_e$ coming within one-pressure scale height 
of the disk mid-plane. Though our calculations show the outflow to be in near
hydrostatic equilibrium near the disk, a two-dimensional calculation of the
disk structure and composition would give a clearer picture of how 
$Y_e$ evolves in outflows from the disk. The uncertainty in $Y_e$
is unfortunate because efficient $^{56}{\rm Ni}$ synthesis hinges
sensitively on $Y_e$ being larger than $0.5$. For the $\alpha=0.03$
disk, the asymptotic $Y_e$ is less sensitive to the vertical disk
structure (at least in our simple calculations) because most of the
change in $Y_e$ occurs a few pressure scale-heights above the disk
mid-plane. 

\subsubsection{The Dynamical Timescale}

Both the final entropy and electron fraction are set by processes near
the disk. By contrast, the dynamic timescale at the epoch of
nucleosynthesis is determined by the wind structure at $r\sim 2-5\times
10^8\cm$.  One way to estimate the timescale characterizing the
expansion of the fluid at $T\lesssim 1/2 {\rm MeV}$ is simply to use
the calculated wind profiles. This is likely not correct. It seems
implausible to expect that the disk outflow will remain well
collimated in quasi-cylindrical geometry for $z\gtrsim 10r_0$. More
likely is that the magnetic or pressure confinement breaks down at large radii
and the wind assumes a quasi-spherical expansion and begins to coast. 
A further dynamical affect may also influence $\tau_{\rm dyn}$
at large radius: when outflow encounters the overlaying stellar 
mantle, which also happens in NS winds, the wind will be  slowed.

For lack of a calculation of the interaction between the disk wind
and the exploding star, we make the assumption that after the sonic point
the wind begins expansion with $a\propto r^2$ (here $a$ is the area defined
in eq.~\ref{aequation}). While the wind has appreciable enthalpy
the expansion is roughly homologous with $v\propto r$ and a 
dynamic timescale $\tau_{\rm dyn}=r/v$. Entropy and mass conservation
imply the scaling $T\propto \rho^{1/3} \propto 1/r$. Once the velocity 
asymptotes to $v_{\rm f}$ a coasting phase described by $\rho
v_{\rm f} r^2={\rm const.}$ follows. In the coasting phase the dynamic
timescale is again approximately $r/v_{\rm f}$, though it is now an increasing
function of radius. In Table \ref{tbl2} we list the dynamic timescales
calculated as described here.The first $\tau_{\rm dynamic}$ listed
for each calculation is that appropriate for homologous expansion ($r_{\rm sonic}
/v_{\rm sonic})$. The second
dynamic timescale listed is that for the flow when it is coasting. For consistency with 
the first definition of $\tau_{\rm dyn}$ this dynamic timescale is defined
as the time needed for the temperature to decrease by a factor of $e$ from 
$T_9\equiv T/10^{9}K=5$. The two different dynamic timescales should 
approximately bracket the plausible range of dynamic timescales. 

Determining the dynamic timescale is equivalent to determining the mass outflow
rate
\beq
\label{mdotsph}
\dot M^{\rm (sph)}\equiv4\pi r^2 \rho v_{\rm f}
\eeq
for a given asymptotic entropy. Here $\dot M^{\rm (sph)}$ is the mass
outflow rate that would obtain if the outflow were spherical. For
winds from the inner regions of accretion disks the true mass outflow
rate is typically much smaller that $\dot M^{\rm (sph)}$ because of
the collimation of the wind. Observations of $^{56}{\rm Ni}$ from GRBs
may help constrain $\dot M^{\rm (sph)}$ (see below). Taking $r\approx v_{\rm f} t$
gives the time at which a given temperature is reached in the outflow
\beq
\label{tofT}
t=0.022 \sec \sqrt{ \frac{{\dot M}_{-1} {s_{30}}}{v_{0.1}^3}}
\left(\frac{0.5{\rm MeV}}{T}\right)^{3/2}.
\eeq
Here ${\dot M}_{-1}=\dot M^{\rm (sph)}/0.1\Msunsec$, $s_{30}=s_{rad}/30$, and $v_{0.1}=v/0.1c$.
Equation (\ref{tofT}) determines the coasting dynamic timescale (i.e. the temperature 
e-folding time when $T_9=5$) as
$\tau_{\rm dyn}=0.1\sec \sqrt{{\dot M}_{-1}s_{30}/v_{0.1}^3}$.

At some point the wind from the disk will encounter the overlaying stellar
mantle. This will influence nucleosynthesis in the wind if the wind is slowed
before $r\approx 2\times10^{8}{\cm}\sqrt{{\dot M}_{-1} s_{30}/v_{0.1}}$,
 where $T_9\approx 2.5$. At times less than a few seconds after disk/black hole
formation, then, a simple wind picture is inadequate. For winds leaving 
the disk at times greater than a few seconds after disk formation the
wind/envelope interaction is not important if the wind is energetic. 
Calculations by \cite{macfadyen} 
show that at $t=9.48$ sec the wind has cleared a
region out to $r\gtrsim 3\cdot 10^9\cm$ in the star. Calculations by 
\cite{mae03} of outgoing bi-polar jets - whose influence on the star may 
roughly approximate the influence of the disk wind - show that by 
$t=5$ sec outflow along the rotation axis continues uninhibited out to $r\sim
2.5\cdot 10^{10}\cm$. This is somewhat faster than the shock
velocities typical of ordinary core collapse SNe ($v_{\rm shock}\sim 10^9\cms$).

 Though the wind/envelope interaction is not expected to influence 
nucleosynthesis in the wind, interaction of the wind
with the stellar mantle will ultimately 
slow the outflow and influence the observed
Ni velocity. As a very rough estimate, if $2\Msun$ of Ni-rich ejecta 
mixes with $10\Msun$ of stellar mantle, the observed Ni velocity will be a 
factor of $\sqrt{2/10}\sim 1/2$ smaller than the estimates in Table \ref{tbl1}.

\subsubsection{The influence of nuclear recombination on the wind dynamics}
\label{nucrecomb}

During $\alpha$ recombination the total entropy is constant (apart from 
the influence of external heating sources), but entropy is transferred from the 
nucleons to the $e^{\pm}/\gamma$ plasma. The amount of entropy transferred can be seen 
from eq.~(\ref{sequation}). For the modest entropy outflows discussed here,
$\ln(T_{\rm MeV}^{3/2}/\rho_8)\approx 3$ during recombination. So, the pair plasma
gains about $(12+3)-(1/4)(15.4+3)\approx 10$ units of entropy. An equivalent way 
to estimate the entropy transfer is to note that $\alpha-$particles are bound
by $\sim 7.074{\rm MeV/nucleon}$, so $\Delta s \approx 7.07{\rm MeV}/T_{\rm rec}\approx 10$,
where $T_{\rm rec}\approx 0.7{\rm MeV}$ is the recombination temperature. $^{56}{\rm Ni}$
is bound by 8.6 MeV/nucleon. This means that synthesis beyond He will add more energy 
to the pair plasma. Our calculations neglect the influence of this extra energy input on 
the wind dynamics. In part, this is because of the difficulty of coupling a nuclear network
to our calculations, and in part because the lion's share of the energy release is from $\alpha$-
recombination (7 out of 8.6 MeV). 

Roughly speaking, nuclear recombination can influence the wind dynamics in two 
different ways. If recombination occurs below the sonic point there is the potential 
for the shift in entropy (as well as pressure and energy) from nucleons to 
the pair plasma to change the amount of viscous and neutrino heating suffered by 
the outgoing wind. This would change the asymptotic entropy, the mass outflow rate, 
and the electron fraction in the wind. Of the models we discuss in this paper, only the
$\alpha=0.03$ disks have outflows with sonic point temperatures below the recombination
temperature. For these we have calculated wind solutions with and without alpha recombination.
The winds without alpha recombination are equivalent to assuming that alpha particles 
are unbound. As can be seen from Tables \ref{tbl1} and \ref{tbl4}, alpha recombination has 
a modest influence on our estimates of the asymptotic entropy and mass outflow rate. This is 
not so surprising, since recombination occurs at relatively low temperatures and after 
most of the heating has occured. 

The second, and most important
effect of recombination is simply to make the nuclear binding
energy available for kinetic energy (eq.~\ref{bern}). Indeed, as \cite{woo03}
have pointed out, the observed kinetic energy for SN2003dh ($\sim 2.6-4\cdot
10^{52}{\rm erg}$) is not so different from the energy liberated by 
recombination of a solar mass of free nucleons 
to Ni ($\sim 1.65\cdot 10^{52}{\rm erg}$). Observationally, though, it still
cannot be determined if Ni recombination accounts for essentially all of 
the observed kinetic energy, or only $\sim 1/3$ of the observed kinetic energy.
Our simple calculations indicate that Ni recombination accounts for about
70$\%$ of the kinetic energy of SN2003dh if the accretion disk has
$\alpha=0.03$, and about 25$\%$ of the kinetic energy if the accretion disk
has $\alpha=0.1$.

\subsection{Outflows Flung Magnetically from the Disk}
\label{flung}

In the above discussion we assumed that the bulk of the work in
driving the outflow is done by pressure/entropy gradients established
by viscous and neutrino heating. It is also possible that material can
be flung outward along a magnetic field line with little or no help
from pressure gradients. 
This case has been thoroughly discussed by
\citet{bla77}, who showed that such outflows might mediate angular
momentum transfer in tenuous accretion disks.  In collapsar
environments it is unclear to what extent such a mechanism can operate
because the outflows are so dense and their inertia is important.

It is worth noting, though, that outflows centrifugally pushed along
a magnetic field line can be qualitatively different from the wind-like
outflows discussed above. In general they are neutron-rich and do not
synthesize $^{56}{\rm Ni}$. The asymptotic entropy and electron fraction 
can both be very low - even lower than the in-disk values if 
neutrino cooling is important. For example, if $\theta_0$ for model D is 
changed to $70\degr$, then the asymptotic entropy and $Y_e$ become $\approx
23$ and $\approx 0.46$ respectively. This can be understood from
the discussion of the Bernoulli integral in section \ref{entsection}.
For $\theta_0=70\degr$ and  $\xi_z=2r_0$, the rotational velocity at $z=2r_0$
is about 50\% larger than the keplerian velocity there, so that
thermal heating does not have to do much work (i.e. $1+U_c$ is small). 
If $\theta_0=70\degr$ and
$\xi_z=1r_0$, so that heating must do the work against gravity, the asymptotic
entropy and electron fraction are essentially the same as for 
$\theta_0=80\degr$ ($s=28,Y_e=0.50$). To summarize, the larger that  
magnetic or other effects keep $r\dot \phi$, the smaller will 
be the final entropy and electron fraction.

\subsection{Nucleosynthesis in Winds}

Here we concentrate on the synthesis of radioactive $^{56}{\rm Ni}$. 
There is no significant production of $r-$process elements for the high
electron fractions and modest entropies found in our calculations. Although material
leaving the disk from very near the hole will have higher entropy,
$Y_e$ will be too large for the $r-$process except perhaps in outflows
with very rapid expansions \citep{mey02}. 

Winds that are accelerating at the epoch of nucleosynthesis ($T\sim0.5$ MeV) generally expand
too quickly for efficient ${\rm Ni}$ production. There is no time for the
3-body reactions that lead to $^{12}{\rm C}$ synthesis and efficient
$\alpha-$captures. In contrast, the evolution and nucleosynthetic yields of winds
that are not accelerating, but coasting, are largely described by two parameters 
(see eq.~\ref{tofT}). For convenience we take these to be the asymptotic 
entropy $s$ and
\beq
\label{beta}
\beta\equiv { \frac{ {\dot M}_{-1}}{v_{0.1}^3}}\propto\tau_{\rm dyn}^2.
\eeq
${\rm Ni}$ synthesis also depends on $Y_e$, though the final ${\rm Ni}$
mass fractions do not vary greatly for $0.5\lesssim Y_e \lesssim 0.53$.

\begin{figure}
\epsscale{0.8}
\plotone{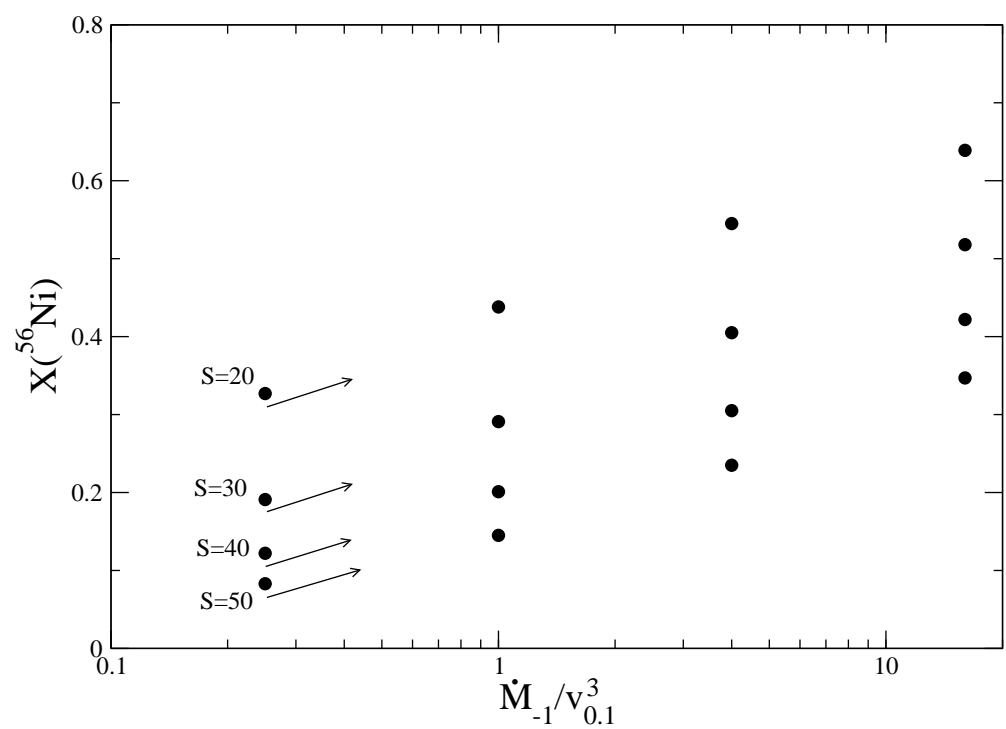}
\caption{Plot of Ni mass fraction vs. the parameter $\beta$
for different entropies. This plot assumes $Y_e=0.51$. \label{nifig}}
\end{figure}

In Figure \ref{nifig} we show final Ni mass fractions as a function of
the entropy and the parameter $\beta$. Large Ni mass fractions are
favored by lower entropies and larger $\beta$'s. At an entropy of 50,
which is obtained in material leaving the disk from $r_0\lesssim 3\times
10^{6}{\rm cm}$, Ni synthesis is inefficient unless
$\dot{M}_{-1}/v_{0.1}^3 \gtrsim 4$. For material with an entropy of
30, which is characteristic of material leaving the disk at $r\sim
10^{7} \cm$, Ni synthesis is efficient ($X_{\rm Ni}>0.25$) as long as
$\dot{M}_{-1}/v_{0.1}^3>1/4$. We note again that the $\dot{M}$ here 
is the isotropic equivelant mass outflow rate. The true outflow rate
is smaller because the Ni-wind is confined to a fraction of the solid 
angle above the disk. 

As far as Ni synthesis in collapsars is concerned, our results are
promising. Table \ref{tbl4} lists the mass outflow rates and final Ni
mass fractions for the different wind calculations. Note that the mass
outflow rates in Table \ref{tbl4} are representative of the ``true''
mass outflow rates, and that $\dot M^{\rm (sph)}$ (defined through
eq.~\ref{mdotsph}) is much larger. The outflows we calculate for
material leaving the disk at $10^{7}{\rm cm}$ all efficiently
synthesize Ni and have rather fast expansion velocities $v\gtrsim 0.1
c$. However, the simple considerations presented here suggest that a
low-alpha disk may have trouble synthesizing $\sim 0.5 \Msun$ of
Ni. The reason is that such a disk is efficiently cooled by neutrino
losses, so that the disk material is tightly gravitationally bound
(see the Bernoulli parameters in Table \ref{tbl3}), and heating cannot
drive large mass outflow rates. For the $\alpha=0.03$ disk, the mass
outflow rates are about an order of magnitude smaller than for the
$\alpha=0.1$ disk. For the $\alpha=0.1$ disk our models predict
outflow rates $4 \pi r_0^2 \rho_0 v_0 \sim 0.07 \Msunsec$ for $r_0=
10^7$ and a factor of about 3 smaller for material leaving the disk
from $r_0=10^{6.5}\cm$. These numbers are in the right range for
explaining a total disk ejecta mass of $\sim 2\Msun$ from an event
with a duration typical of long duration GRBs ($\sim 10-100\sec$).

\section{Impulsive Rapid Mass Ejection: Magnetic Bubbles}
\label{section:bubble}

We imagine a background disk wind, similar to the solutions obtained
in the preceding section, but modulated by rapid impulsive events 
due to magnetic reconnection.  Highly magnetic (low $\eta=P/(B^2/8\pi)$)
filaments may be formed in the disk mid-plane and emerge into the steady
wind background rapidly as they expand and accelerate in the approximately 
exponential atmosphere of the disk.  We propose here that this environment will
have modest entropy, but
low electron fraction (similar to the $Y_e$ that obtains at the disk mid-plane)
on account of the rapid expansion of the bubbles.  It is in these events that 
$A\sim 130$ $r$-process elements may be synthesized.

There is an important difference between heating from magnetic
instabilities and heating in an alpha-disk model. The viscous heating rate
in an alpha-disk tracks the density. In eq.~(\ref{sdot}), for example, it is
seen that $\dot s\sim P/(\rho T)\sim {\rm
Constant}$, so that very rapid increases in entropy can not be obtained
in such a model. By contrast, heating from magnetic instabilities depends
on the geometry of the magnetic fields. There is nothing to prevent 
heating in a relatively baryon dilute region (say two pressure scale
heights above the disk mid-plane). Such heating can result in rapid, large
increases in the entropy of the outflowing material. This basic idea is 
behind a number of suggestions for the origin of relativistic outflow
in GRBs \citep{narayan_pac,klu98}. 

To see how bubbles might form in the collapsar accretion disk, consider the
following argument adapted from \citet{klu98}, who studied
differentially rotating neutron stars. 
For simplicity, consider an isothermal disk with a structure and
equation of state given by 
\beqa
\label{simpledisk1}
\rho(z)& = & \rho_0\exp(-z) \\
s(z) & = & T^{3/2}/\rho  = s_0\exp(z)\\
P(z) & = &\rho T  =  P_0\exp(-z).
\label{simpledisk3}
\eeqa
Here we have assumed the gas to be dominated by free nucleons in the disk
and have scaled the height above the disk in units of the disk scale
height $H$.
A fluid element rising adiabatically from the mid-plane of the disk will 
be in pressure equilibrium with the background fluid. If there is no 
magnetic pressure, then this element (denoted with the subscript $b$)
 will have a density
\beq
\rho_b=\rho_0\exp (-3z/5)>\rho(z), 
\eeq
which implies that the fluid element will fall back to the mid-plane.
If the fluid element carries a magnetic pressure $P_M$, pressure equilibrium
with the background disk implies 
\beq
\rho_b=\rho_0\exp\left(-3z/5\right)\left(1-\frac{P_M}{P(z)}\right)^{3/5}.
\eeq
If the magnetic pressure is large enough then $\rho_b<\rho(z)$ and the fluid
element will be buoyant. The force per unit volume on the fluid element is
$\Omega_k^2 z$, which gives the equation of motion 
\beq
\label{zddot}
\ddot{z}=z\left(\frac{\rho(z)}{\rho_b}-1\right).
\eeq
Here we have scaled time in units of $\Omega_k^{-1}$. If we make the simple
assumption that the magnetic pressure increases linearly with time 
and $P_M=\eta P_0(t/2\pi)$, then eq.~(\ref{zddot}) becomes 
\beq
\ddot{z}=z\left[\exp\left(-3z/5\right)\left(1-\frac{\eta t}{2\pi}\exp(z)\right)^{-3/5}-1\right].
\eeq
This equation can be solved for $z(t)$ to provide an estimate of 
$P_M(z)$ and the entropy increase of the bubble as a function of the
reconnection height. As a simple approach, suppose that an element
rises one pressure scale height per radian that the disk rotates, or
$z=t$ in our notation. This is a reasonable assumption since the 
magnetic pressure quickly becomes large compared to the disk pressure
as the filament rises. If $z\approx t$, the ratio of magnetic energy 
density to thermal energy density evolves approximately as 
\beq
\label{pmovere}
\frac{P_M}{\rho_b T_b}=\frac{\eta z}{2\pi \exp(-z)-\eta z}.
\eeq
If the energy in magnetic fields is transferred to thermal energy 
of the buoyant bubble the fractional increase in entropy of the bubble
is approximately the ratio given in eq.~(\ref{pmovere}). As an example, suppose
that the initial magnetic field is $10\%$ of the equipartion field 
($\eta=0.1$). If the energy in magnetic fields is transferred to the bubble
at 1 disk-scale height above the mid-plane
the fractional increase in entropy will be $\approx 6\%$. If the energy is 
transferred at three scale heights above the mid-plane, entropy will increase 
by a factor of $\sim 23$.

To assess if the above argument seems plausible, 
let us assume that the magneto-rotational instability (the MRI; 
Balbus \& Hawley 1994; Balbus \& Hawley 1998) operates in collapsar 
disks and that this produces the turbulent viscosity necessary for accretion.
Then, the local saturation magnetic field at a radius $r_0$ will be in
rough equipartition with the azimuthal kinetic energy density;
\beq
\frac{B_{\rm sat}^2(r_0)}{8\pi}\sim\frac{1}{2}\rho v_\phi^2
\label{saturation1}
\eeq
where $v_\phi=r_0\Omega_k=r_0(GM/r_0^3)^{1/2}$.  Empirically, for simulations
of the MRI in accretion disks (Hawley, Gammie, \& Balbus 1996)
the magnetic field saturates at a
sub-equipartition value, typically $1/2\pi$ times the field estimated 
from
eq.~(\ref{saturation1}).  Taking this into account,
\beq
B_{\rm sat}(r_0)\sim1\times10^{14}\,\,{\rm
G}\,\,\rho_9^{1/2}\,M_3^{1/2}\,r_{0_7}^{-1/2},
\label{saturation2}
\eeq
where $\rho_9=\rho/10^{9}$ g cm$^{-3}$, $r_{0_7}=r_0/10^7$ cm and $M_3=M/3\Msun$. 
This magnetic
field strength is in the right range for explaining modest entropy increases
in the disk material. 
If the MRI is operating, the time to build a magnetic field is set by
$\Omega_k^{-1}$.
The timescale for the maximum growing mode is
\beq
\tau_{\rm Max}=4\pi\left|\frac{d\Omega_k}{d\ln r}\right|^{-1}.
\eeq
Roughly, we may assume that the local magnetic field may be entirely
built and dissipated in $\tau_{\rm Max}$.
Using eq.~(\ref{saturation1}), and estimating the reconnection time as
$\tau_{\,\rm Rec}\approx \tau_{\rm Max}=L_{\,\rm Rec}/v_{\rm A}$,
where $v_{\rm A}$ is the Alfv\'{e}n speed, we find that the
characteristic length scale for reconnections $L_{\rm \,Rec}\sim
115\,\,{\rm km}$.  This may seem uncomfortably large, but we note that
the $\tau_{\rm \,Rec}$ above is the minimum required -- we should
perhaps require 3-4 doublings of the field at $r_0$ before
dissipating.  In this case, the reconnection time would be longer and $L_{\rm \,Rec}$ 
would be commensurately smaller.  We expect these reconnection events to happen
impulsively, and be separated in time by $\sim\tau_{\rm \, Rec}$.

A bubble will begin to rise after entropy is deposited in a localized 
region. If the bubble is formed in a region of the disk where approximate
hydrostatic equilibrium obtains, the bubble height $z$ will evolve as
\beq
\label{zindisk}
z\approx z_0 \Omega_k t.
\eeq
Here we have made use of the fact that the gravitational force in the
vertical direction is approximately $\Omega_k^2 z$. The characteristic length
scale $z_0$ is taken here to be the pressure scale height $H$ in the disk.
The rising bubble will be in pressure equilibrium with the background
fluid. This implies that initially the temperature evolves according
to 
\beq
\label{Tindisk}
T\propto \exp(-z/4H)
\eeq
provided that the bubble is radiation dominated. These considerations suggest
that the timescale characterizing the initial expansion of the bubble
should be of order $\tau_{\rm initial}\approx 10/\Omega_k\approx 0.01 \sec$.
Noting that the positron capture rate is $\lambda_{e^+{\rm n}}\approx
11 {\sec^{-1}}(T/3{\rm MeV})^5$ gives an estimate of the change in $Y_e$
of the outgoing bubble
\beq
\label{deltaye}
\Delta Y_e\approx 0.2\left( \frac{T_0}{3\rm MeV}\right)^5 \left(
\frac{\tau_{\rm initial}}{0.01\sec} \right).  
\eeq 
Here $T_0$ is the
temperature of the bubble formed after magnetic reconnection. We note
that for the $\alpha=0.03,\dot M=0.1\Msunsec$ disk, the height
averaged disk temperature is $\sim 2.2-2.8{\rm MeV}$ for
$r_0<10^7\cm$. Bubbles can retain low values of $Y_e$.

Equations (\ref{zindisk}) and (\ref{Tindisk}) are appropriate for
characterizing the bubble passage through the region of the disk that
is in approximate hydrostatic equilibrium. Those equations are not
appropriate for estimating the expansion timescale at the late times
and low temperatures important for nucleosynthesis. The expansion
timescale at $T\sim 0.5{\rm MeV}$ depends on the profile of the
background wind. If the background wind is radiation dominated well
above the disk, then pressure equilibrium between the bubble and wind
implies that the temperatures in the wind and bubble are equal. In
this case the expansion timescale of the wind ($\tau$ in Table 2) may
be used as a rough indication of the expansion timescale of the
bubble.

\subsection{Nucleosynthesis in Bubbles}

As discussed in the introduction, there are indications that the site
of the $A\lesssim 130$ $r$-process nuclei is different from the site of
the $r$-process nuclei with $A\gtrsim 130$. In particular, observations
of ultra-metal poor stars (e.g.~Sneden et al.~1996; Burris et al.~2000)
and inferences from elemental abundances in
pre-solar meteorites (Wasserburg, Busso, \& Gallino 1996)
suggest that the ${\rm ^{135,137}Ba}$ isotopes are not
significantly produced in the same events that produce $^{127,129}{\rm
I}$ and lighter $r$-process elements. An exhaustive survey of how 
a nuclear burning site can produce a near solar abundance pattern
of nuclei near one of the $r$-process peaks, while leaving the other
peaks unpopulated, is beyond the scope of this paper. Here we show 
in broad stroke that conditions in collapsars are favorable for the
synthesis of the $A\sim 130$ $r$-process elements. In the spirit of 
investigation, we also discuss a way in which a near-solar 
abundance pattern of (only) $A\leq 130$ $r$-process elements can 
be synthesized.

\begin{figure}
\epsscale{0.8}
\plotone{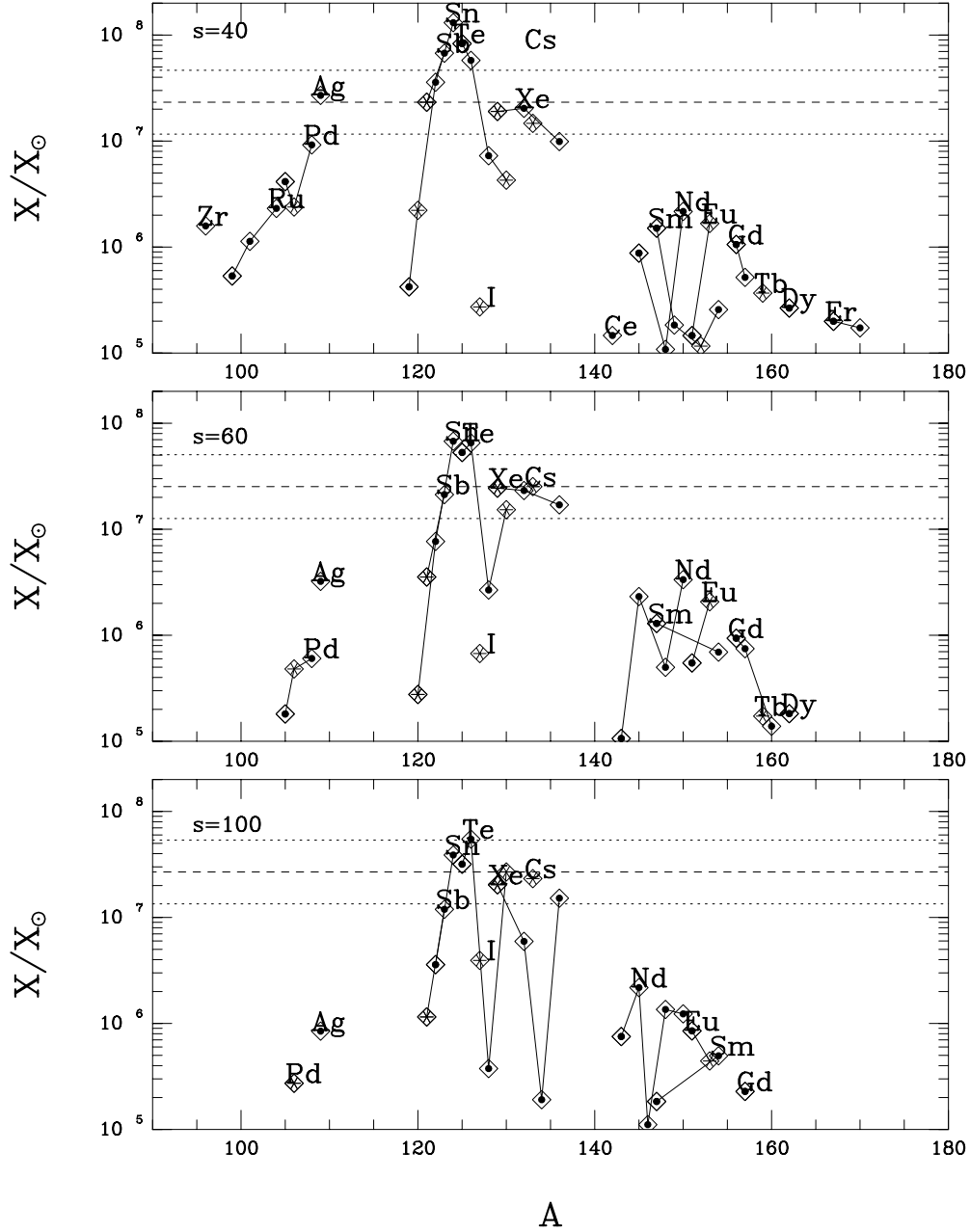}
\caption{Overproduction factors for nuclei synthesized
in bubbles with $Y_e=0.2$, $\tau=0.12\sec$, and different
entropies. Solid lines connect isotopes of a given element.
The most abundant isotope
in the Sun for a given element is plotted as an asterisk. A diamond
around a data point indicates the production of that isotope as a
radioactive progenitor.
Though some $r$-process elements are synthesized, 
agreement with the solar abundance pattern is poor.\label{badrfig}}
\end{figure}

Neutron-rich outflows characterized by rapid expansion generically
synthesize $r$-process elements (e.g. \citet{hof97}). As an example, 
we show in Fig. \ref{badrfig} results of nucleosynthesis calculations
for outflows with $Y_e=0.2$, a temperature e-folding time 
$\tau=0.12\sec$, and three different entropies. These conditions are close to 
those we estimate may obtain in outflows from the inner regions of 
accretion disks with $\alpha=0.03$ and $\dot M=0.1\Msunsec$. Here the 
(unnormalized) overproduction factor for nucleus $j$ is defined as
\beq
O(j)=\frac{X_j}{X_{\odot,j}},
\eeq
where $X_j$ is the mass fraction of the nucleus $j$ in the bubble
and $X_{\odot,j}$ is the mass fraction of the nucleus in the sun.
Though it is evident from Fig. \ref{badrfig} that some $r$-process
elements are synthesized, good agreement with the solar abundance
pattern is not obtained. 

It is unreasonable to expect that a single type of bubble will prevail
in the dynamic collapsar environment. Rather, a broad spectrum of bubbles,
with different entropies, dynamic timescales, electron fractions, and so 
on will be created. To investigate the average nucleosynthesis we 
generated random bubbles with properties defined according to 
\begin{eqnarray}
\label{bubbleprod1} s &=& 50+50r \\
Y_e &=& 0.15+0.25r \\
\tau &=& 0.03(1+4r)\sec \\
T_{9,{\rm mix}} &=& 1+2r.
\label{bubbleprod4}
\end{eqnarray}
Here $r$ is a random number between 0 and 1 and is generated separately for
each of eqs.~(\ref{bubbleprod1}-\ref{bubbleprod4}). The choices for $Y_e$
and $\tau$ above were adopted
because they represent the expected range of conditions in high entropy
outflows from the inner regions ($r\lesssim 10^7\cm$) of a disk with 
 $\alpha=0.03$ and $\dot M=0.1\Msunsec$. Choices for $s$ and 
$T_{9,{\rm mix}}$ are not well constrained. The parameter $T_{9,{\rm mix}}$
is the temperature at which the bubble is assumed to mix with the 
proton-rich ambient medium. When the bubble shears or destabilizes,
the free neutrons available for capture are diluted. To represent this
we made the rough approximation that all neutron captures then cease
and the abundance pattern of neutron rich elements is frozen in (except for
$\beta$ decays) for $T_9<T_{9,{\rm mix}}$. Because charged particle reactions
are relatively slow for $T_9\lesssim 3$ we expect this to be a fair 
approximation. 

\begin{figure}
\includegraphics[scale=0.6,angle=270]{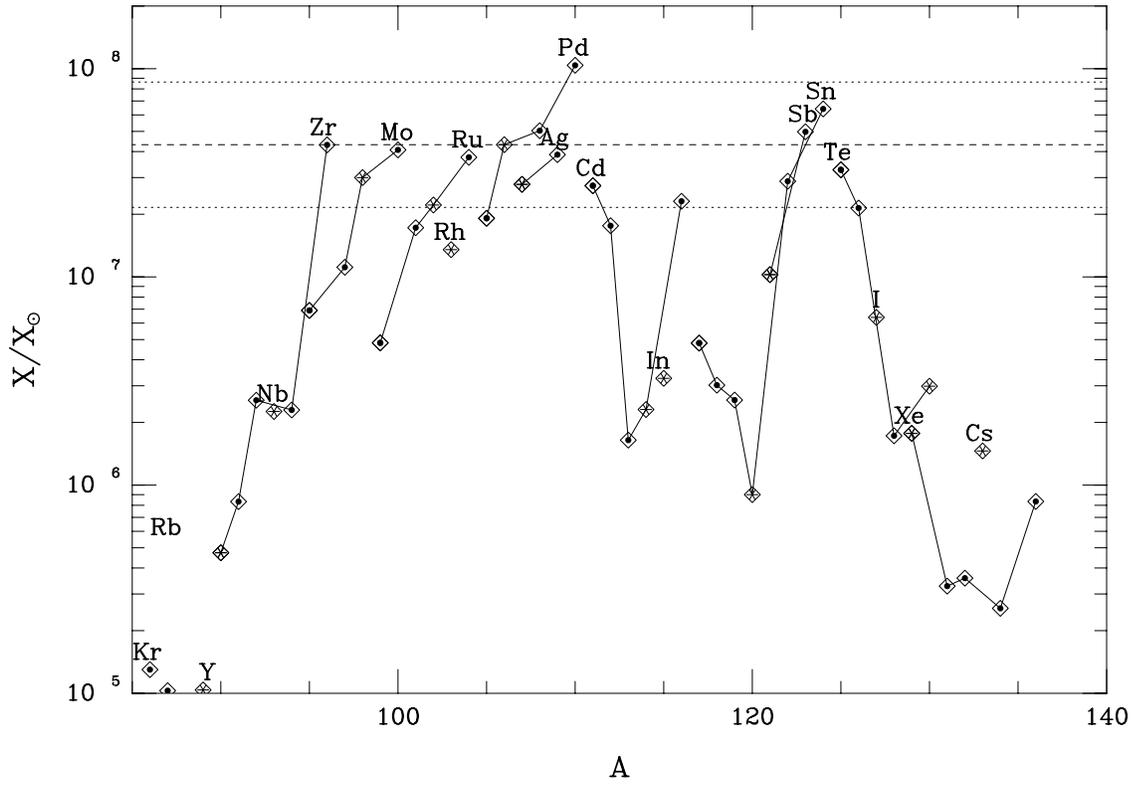}
\caption{Average overproduction factors for 100 bubbles 
generated according to eqs.~(\ref{bubbleprod1}-\ref{bubbleprod4}). 
Agreement with the
solar abundance pattern of $r$-process elements with $A<130$ 
is quite good, though $^{127}{\rm I}$ is underproduced by a factor of
about 4. Production of species heavier than $A=130$ is negligible. \label{combine}}
\end{figure}

\begin{figure}
\plotone{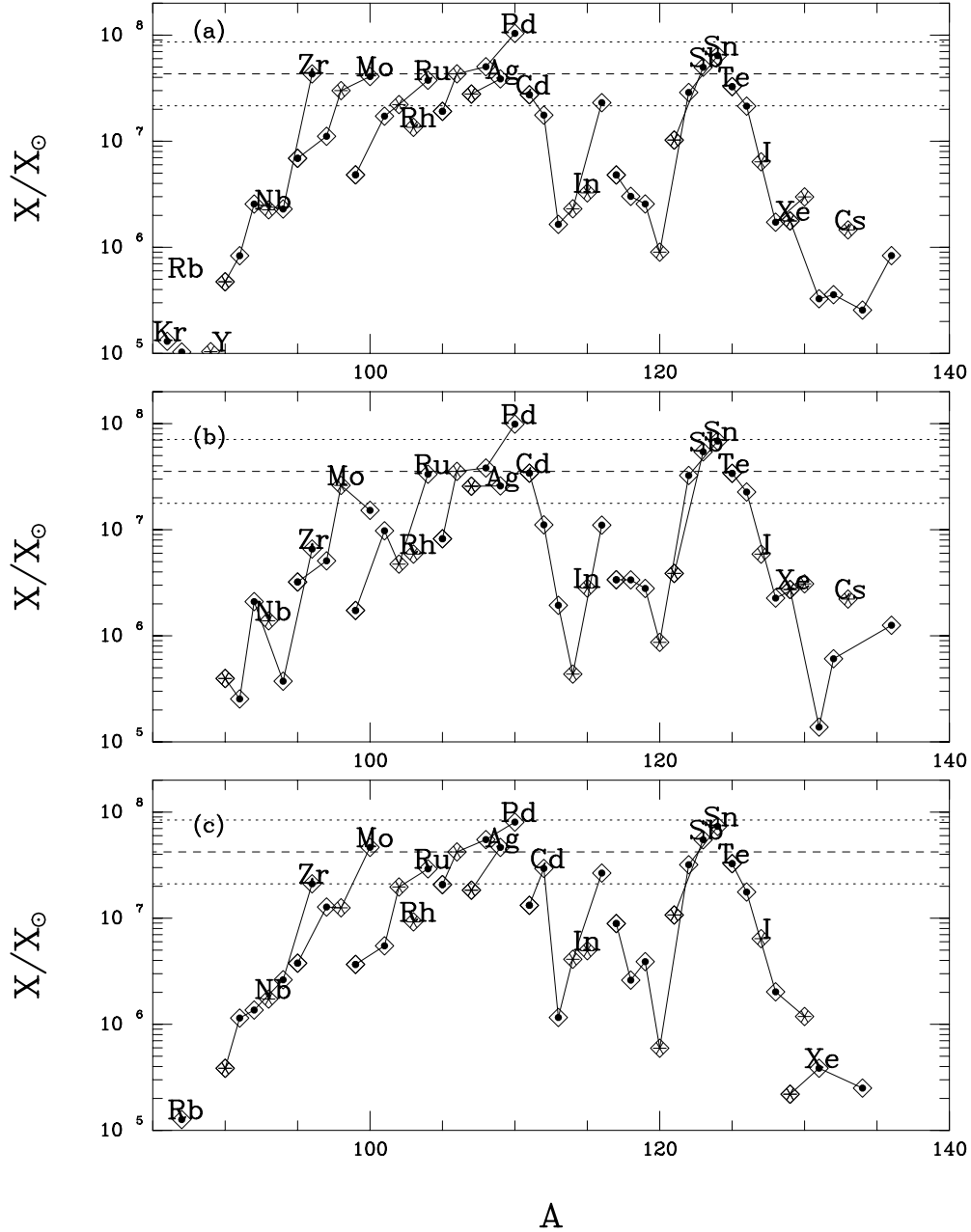}
\caption{Influence of changing assumptions about the
distribution of bubbles produced. Run (a) is the reference run and corresponds
to the distribution of bubbles described by 
eq.~(\ref{bubbleprod1}-\ref{bubbleprod4}). Run (b)
is the same as run (a) except the entropy in the bubbles is assumed to 
be uniformly spread from $s=50$ to $s=150$. Run (c) is the same as run (a),
except $T_{9,{\rm min}}$ is assumed to be uniformly spread from 1.5 to 3.5.
In run (a) the average of 100 bubbles was taken while in runs (b) and (c) 
the average of 30 bubbles was taken.
\label{3randoms}}
\end{figure}

Average overproduction factors for one hundred bubbles generated as
described above are shown in Fig. \ref{combine}. Overall there is
quite good agreement with the solar abundance pattern of $90<A<130$
$r$-process elements. Overproduction factors for $^{100}{\rm
Mo},^{107}{\rm Ag},^{110}{\rm Pd},^{116}{\rm Cd},^{123}{\rm Sb}$ and
$^{124}{\rm Sn}$ - elements principally or entirely synthesized in the
$r$-process - are all within a factor of a few of the overproduction
factor for $^{106}{\rm Pd}$.  $^{127}{\rm I}$ is underproduced by a
factor of about 5.  Most of the $A\sim130$ nuclei have progenitors
about 8 units from stability near the N=82 closed neutron shell in our
calculations. Uncertainties in the location of the shell closures and
the $\beta$ decay rates of near-drip line nuclei may account for the
modest underproduction of $^{127}{\rm I}$.  $^{135}{\rm Ba}$ and
heavier elements are absent. In Fig. \ref{3randoms} we investigate the
sensitivity of nucleosynthesis to the distribution of $s$ and
$T_{9,{\rm min}}$ in the bubbles. Nucleosynthesis is relatively
insensitive to modest changes in
eqs.~(\ref{bubbleprod1}-\ref{bubbleprod4}).

Our calculations indicate that typical overproduction factors for 2nd-peak 
nuclei are $\bar O \approx 5\cdot 10^7$. With this, we can estimate how
much mass collapsars have to eject as bubbles in order to account for
the galactic inventory of $A<130$ $r$-process nuclei. For events with the 
frequency of Type II SNe, studies of galactic chemical evolution 
indicate that a typical nuclide must have an overproduction factor
\beq
\label{oeqten}
O(j)\approx 10 \frac{\bar M^{ej}}{M^{b}}
\eeq
in order to account for the observed solar abundance of that 
nuclide (e.g. \citet{mat92}). 
In eq.~(\ref{oeqten}) $\bar M^{ej}\approx 10-20 \Msun$ 
is the typical mass ejected by a type II SN and $M^b X(j)$ is the 
mass of the nuclide $j$ ejected. For collapsars, $O(j)$ must be 
$\sim 1/f_c$ larger, where $f_c$ is the fraction of core collapse
SNe that become collapsars. The total mass ejected as bubbles must then
\beq
\label{massneeded}
M^b\approx 10^{-4}\Msun \frac{0.1}{f_c} \frac{\bar O}{5\cdot 10^7}.
\eeq
If typical
bubbles are formed with initial radius $r_b$, temperature $T_b$, and entropy
$s_b$ per baryon, eq.~(\ref{massneeded}) implies that the number of bubbles
needed per event is 
\beq
\label{nbubbles}
n_b\approx 500 \frac{s_b}{50} \left(\frac{10^6\cm}{r_b}\right)^3 \left(\frac{{2\rm MeV}}{T_b}\right)^3 \frac{M_b}{10^{-4}\Msun}.
\eeq
Now, if the disk lasts for a time $t$, the number of disk 
revolutions, or magnetic field windings, per bubble is 
\beq
n_{\rm wind}=\frac{t\Omega_k}{2\pi n_b}\approx 8 \frac{t}{50\sec} \frac{\Omega_k}{10^3 {\rm sec}^{-1}} \frac{10^3}{n_b},
\eeq
which is a reasonable number if magnetic instabilities take a few rotations to
develop. 

We have argued that if collapsars are only $1\%$ as frequent as Type
II SNe then each collapsar needs to eject $10^{-3}\Msun$ of r-process
bubbles in order to account for the total mass of 130-peak r-process
material present in the galaxy today. This required ejecta mass is
approximately $0.01\%$ of the total mass ejected in a Ni wind and is
roughly consistent with a description of bubbles as forming on a
magnetic instability timescale and observed GRB durations. However,
there still remains the more delicate issue of whether collapsars as
the 130-peak site is consistent with observations of metal poor stars
and what is known about galactic chemical evolution. 

Unfortunately, much less is known about the production of the A$<$130
elements than is known about the production of the heavier r-process
elements. For example, while CS22892-052 and a handful of similar
stars are thought to have been enriched by only one $A>130$ production
event, these authors are not aware of any stars thought to have been
enriched by only a single $A<130$ enriching event. Within the context
of the models developed by Qian and Wasserburg this has an explanation
in the relative rarity of the events which produce the second peak
elements.

One type of argument which has been used to differentiate proposed $r$-process
sites is based on an analysis of the refreshment rate of material in the
interstellar medium (ISM). This argument has previously been used to assesss
the viability of neutron star mergers as an $r$-process site and can also be
applied to collapsars. \cite{qia01} showed that enrichment of the ISM with
second peak elements every $10^8{\rm yr}$ is consistent with the observed
trend of ${\rm Ag}$ abundances with ${\rm [Fe/H]}$ in metal poor stars. 
Here we simply note that a refreshment frequency of $f_{\rm ref}\approx
(10^8{\rm yr})^{-1}$ is consistent with enrichment of the ISM by collapsars.
Following the notation of \cite{qia01},
\beq
f_{\rm ref}\approx (10^8{\rm yr})^{-1} 
\left(\frac{f^{\rm SN}}{(30 {\rm yr})^{-1}}\right) \left(\frac{f_c}{0.01}\right)
\left(\frac{M_{\rm mix}}{3\cdot 10^5 \Msun}\right),
\eeq
where $f^{\rm SN}$ is the rate of type II SNe in the galaxy and $M_{\rm mix}$
is the total mass swept up by a collapsar remnant. Note that $M_{\rm mix}$
for collapsars is about an order of magnitude larger than the mass swept 
up by type II SNe. This is because the kinetic energy in the collapsar explosion
is about ten times larger and because the swept up mass is proportional to
$E_{\rm kinetic}^{6/7}$ \citep{thorn}.

\section{Results and Conclusions}

We have considered the nucleosynthesis which may attend outflows
in the collapsar Gamma Ray Burst environment. Both wind-like
outflows and bubble-like outflows were considered.
Wind-like outflows may be relevant for recent observations of
SN1998bw and SN2003dh, which hint at a robust connection between the
central engines of GRBs and core-collapse SNe -- or at least SN-like
light curves. To power such light curves requires $\sim 0.5\Msun$ of 
radioactive Ni moving 
outward very rapidly ($v\gtrsim 0.1c$ for SN2003dh). These are 
characteristics beyond the reach of canonical SNe with energies 
$\sim 10^{51}\erg$. The results of our simple 
models of viscosity and neutrino-driven winds are promising. Under a broad
range of conditions such winds copiously produce fast moving
radioactive Ni.

In general, winds from collapsar disks cannot preserve a large neutron
excess.  This implies that these winds will not synthesize interesting
neutron rich elements. However, chaotic heating or buoyant magnetic
filaments in localized regions in the disk result in bubbles which
rise on a timescale comparable to a Kepler period. This is fast enough
to preserve the neutron excess found in the mid-plane of the
disk. Though we do not have a complete theory of bubble production, we
have shown that the solar abundance pattern for $90<A<130$ $r$-process
elements can naturally be produced in bubble-like outflows and that the 
requirements for the total mass ejected are plausible. As we have
discussed, identification of collapsars as the source of the 2nd peak
$r$-process elements is consistent with a number of observational
indications.

\acknowledgments 

We gratefully acknowledge Stan Woosley for guidance and many helpful suggestions.
JP also acknowledges George Fuller for insightful comments. 
This research has been supported through a grant 
from the DOE Program
for Scientific Discovery through Advanced Computing (SciDAC;
DE-FC02-01ER41176). This work was performed under the
auspices of the U.S. Department of Energy by University of California
Lawrence Livermore Laboratory under contract W-7405-ENG-48.
TAT gratefully acknowledges conversations with Eliot Quataert and
is supported by NASA through Hubble Fellowship
grant \#HST-HF-01157.01-A awarded by the Space Telescope Science
Institute, which is operated by the Association of Universities for Research in Astronomy,
Inc., for NASA, under contract NAS 5-26555.

\clearpage

\clearpage

\clearpage

\clearpage

\clearpage

\clearpage

\clearpage

\begin{deluxetable}{ccccccccc}
\tablecaption{Wind characteristics for different parameters\label{tbl1}}
\tablewidth{0pt}
\tablehead{
\colhead{Model} &
\colhead{$\alpha$} &
\colhead{$r_0(\cm)$} &
\colhead{$\xi_z/r_0$} &
\colhead{$Y_{e,i}$} &
\colhead{$Y_{e,f}$} &
\colhead{$s_f$} &
\colhead{$v_{\rm f}/c$\tablenotemark{2}}
}
\startdata
A & 0.03 & $10^{6.5}$  & 2    & 0.12 & 0.56 & 54 & 0.21[0.17]\\
B & 0.03 & $10^{7}$    & 2    & 0.22 & 0.50\tablenotemark{1} & 24\tablenotemark{1} & 0.11[0.06]\\
C & 0.1  & $10^{6.5}$  & 2    & 0.44 & 0.54 & 52 & 0.36[0.34]\\ 
D & 0.1  & $10^{7}$    & 2    & 0.43 & 0.50 & 33 & 0.23[0.19]\\
E & 0.1  & $10^{6.5}$  & 4    & 0.44 & 0.50 & 47 & 0.32[0.30]\\ 
F & 0.1  & $10^{7}$    & 4    & 0.43 & 0.46 & 32 & \,\,\,0.22[0.18]
\enddata
\tablenotetext{1}{For model B $s_f=26$ and $Y_{e,f}=0.53$ if 
alpha recombination is neglected. For all other models 
alpha recombination does not effect $s_f$ or $Y_{e,f}$ in 
our calculations.}
\tablenotetext{2}{Values in square brackets are those calculated if
alpha recombination is neglected.}
\end{deluxetable}

\clearpage

\begin{deluxetable}{cccc}
\tablecaption{Dynamic timescales for the different winds.\label{tbl2}}
\tablewidth{0pt}
\tablehead{
\colhead{Model} &
\colhead{$\tau_{\rm homologous}(\sec)$} &
\colhead{$\tau_{\rm coast}(\sec)$}
}
\startdata

A & 0.03 &  (NC)\tablenotemark{a}    \\
B & 0.17 &  (NC)\tablenotemark{a}    \\
C & 0.007 & $\sim 0.04$     \\ 
D & 0.03 &  $\sim 0.10$ \\
E & 0.002 & $\sim 0.03$  \\ 
F & 0.012 & $\sim 0.08$ 
\enddata
\tablenotetext{a}{The sonic-point temperature for these winds
is less than $0.5{\rm MeV}$.}
\end{deluxetable}

\clearpage

\begin{deluxetable}{ccccccc}
\tablecaption{Minimum values of the final entropy for a given
	final velocity\label{tbl3}}
\tablewidth{0pt}
\tablehead{
\colhead{$\alpha$} &
\colhead{$r_0(\cm)$} &
\colhead{$b_i({\rm MeV})$\tablenotemark{a}} &
\colhead{$s_f(v_{\rm final}=0.1c)$} &
\colhead{$s_f(v_{\rm final}=0.2c)$} &
\colhead{$s_f(v_{\rm final}=0.3c)$}
}
\startdata

 0.03 & $10^{6.5}$  & $-$132    & 53 & 58  & 66   \\
 0.03 & $10^{7}$    & $-$37    & 26 & 33  & 44   \\
 0.1  & $10^{6.5}$  & $-$83    & 45 & 49  & 56   \\ 
 0.1  & $10^{7}$    & $-$21    & 28 & 35 & 46 
\enddata
\tablenotetext{a}{In-disk value of the Bernoulli parameter (eq.~\ref{bern}).}
\end{deluxetable}

\clearpage

\begin{deluxetable}{ccc}
\tablecaption{Mass ablation rate and estimated final 
$^{56}{\rm Ni}$ mass fractions for some different wind
calculations.\label{tbl4}}
\tablewidth{0pt}
\tablehead{
\colhead{Model} &
\colhead{$\dot M (\Msunsec)$\tablenotemark{a}} &
\colhead{$X(^{56}{\rm Ni})$\tablenotemark{b}}
}
\startdata

A & $10^{-3}$ & $\lesssim0.1$ \\
B & $4\times 10^{-3}$ & $\sim0.5$ \\
C & $3\times 10^{-2}$ & $\sim0.1$ \\
D & $7\times 10^{-2}$ & $\sim0.4$ \\
E & $6\times 10^{-2}$ & $\lesssim0.1$ \\ 
F & $0.14$ & -\tablenotemark{c}
\enddata
\tablenotetext{a}{Defined as $4\pi r_0^2 \rho_0 v_0$.}
\tablenotetext{b}{Estimated from Figure \ref{nifig} under the assumption
that the wind begins quasi-spherical expansion after the sonic point. Dynamic
timescales were taken from Table \ref{tbl2}.}
\tablenotetext{c}{This model has $Y_e<0.48$ and does not produce 
$^{56}{\rm Ni}$.}
\end{deluxetable}

\end{document}